\documentclass[12pt]{article}

\usepackage[utf8]{inputenc}
\usepackage{amsmath,amssymb}
\usepackage{graphicx}
\usepackage[utf8]{inputenc}
\usepackage[english]{babel}
\usepackage{enumerate} 
\usepackage{amsthm}
\usepackage{multirow}
\usepackage{algorithm}
\usepackage{algpseudocode}
\usepackage{algorithmicx}
\usepackage{diagbox}

\theoremstyle{definition}
\newtheorem{definition}{Definition}[section]
\newtheorem{theorem}{Theorem}[section]
\mathchardef\mhyphen="2D
\newcommand{\thedgerel}{\ensuremath{\sim_{e_{\theta}}}}
\newcommand{\graphuni}{\ensuremath{\mathcal{G}}}
\newcommand{\reals}{\ensuremath{\mathbb{R}}}
\newcommand{\sensitivity}[2]{\ensuremath{\Delta_{#1}^{#2}}}
\newcommand{\locsens}[3]{\ensuremath{LS_{#1}^{#2}(#3)}}

\title{Analysis of centrality measures under differential privacy models}

\author{Jesse Laeuchli$^1$, Yunior Ram\'{i}rez-Cruz$^2$\\ 
and Rolando Trujillo-Rasua$^1$\\ 
{\small $^1$Centre for Cyber Security Research and Innovation}\\ 
{\small School of Information Technology, Deakin University}\\ 
{\small 221 Burwood Hwy., Burwood VIC 3125, Australia}\\
{\small $^2$Interdisciplinary Centre for Security, Reliability and Trust}\\ 
{\small University of Luxembourg}\\ 
{\small 6, av. de la Fonte, L-4364 Esch-sur-Alzette, Luxembourg}\\  
{\small \texttt{\{jesse.laeuchli, rolando.trujillo\}\@@deakin.edu.au},}\\ 
{\small \texttt{yunior.ramirez\@@uni.lu}}}

\begin{document}

\maketitle

	\begin{abstract}
	This paper provides the first analysis of the differentially private computation 
	of three centrality measures, namely eigenvector, Laplacian and closeness 
	centralities, on arbitrary weighted graphs, using the smooth sensitivity approach. 
	We do so by finding lower bounds on the amounts of noise 
	that a randomised algorithm needs to add in order to make the output 
	of 	each measure differentially private. Our results indicate 
	that these computations are either infeasible, in the sense that 
	there are large families of graphs for which smooth sensitivity is unbounded; 
	or impractical, in the sense that even for the cases where smooth sensitivity 
	is bounded, the required amounts of noise result in unacceptably large 
	utility losses. 
	\end{abstract}
	

\section{Introduction}

Online social networks are pervasive nowadays. 
Social interactions via digital 
means, such as chat apps, electronic mail, discussion forums and social 
networking websites, are typically chosen over traditional face-to-face 
interactions for various reasons. It could be that digital 
interaction is convenient and efficient, or that it is simply the only 
available communication channel, as in the current COVID-19 crisis. In either 
case, the result is a massive amount of digital information for the public and 
private sectors alike to perform social analysis and improve or optimise their 
services. 

Typical social network analysis tasks are, among others, community detection,
which allows to detect groups of users that display a common behaviour or are 
somehow strongly interrelated; link prediction, commonly used for suggesting 
new friends; and centrality analysis, which helps to determine the role a user 
plays in the network. The challenge is that, allowing accurate analyses on a 
dataset while, at the same time, protecting 
users' private information are conflicting goals~\cite{DR2014}. 
Even aggregated information, such as election results, has the potential 
to unwillingly reveal sensitive information, such as people's vote. 
An extreme, yet illustrative example is an election result 
where a candidate gets all possible votes. 
That reveals, not only 
the preference of the population, but also the choice that each citizen made at 
an individual level. 
One may argue that when an attribute is shared by many it becomes less 
sensitive. 
However, this is likely not the case for attributes such as chronic diseases, 
even though they are suffered by a large part of the population. 

What information needs protection and how information may be leaked are 
fundamental questions that privacy notions, models and frameworks have 
attempted to answer. In general, answers to those questions vary 
depending on the type 
of information that is being collected and analysed. Voters may want 
to keep their votes secret, users in a social network may like to 
protect their relationships with others, etc. This explains the explosion of 
privacy models that can be found in literature; each addressing a specific 
scenario; each unable to claim privacy outside of its application domain. 

In 2006, Dwork, McSherry, Nissim and Smith formulated privacy in a different 
way~\cite{Dwork2006}. They addressed the question of how a data holder can 
promise to each user that the analysis results will be independent (up to some 
extent) of their contribution to the dataset. Rather than defining 
what information needs protection, Dwork et al. aimed at ensuring 
that the fact that an individual contributed her data to a survey is unlikely 
to be determined from the output of the analyses, regardless of what the survey 
is about. The resulting privacy definition, called \emph{differential privacy} (DP), 
is formalised as follows. 

\begin{definition}[$(\varepsilon, \delta)$-differential privacy]
\label{def-diff-privacy} 
	Let $\mathcal{D}$ be a universe of datasets and 
	$\sim$ a symmetric and anti-reflexive \emph{neighbouring} relation on 
	$\mathcal{D}$ relating two datasets if one can be obtained from the other 
	by adding or removing one entry. A randomised algorithm 
	$\mathcal{A}:\mathcal{D}\rightarrow\mathcal{O}$, for some co-domain $\mathcal{O}$,
	is $(\varepsilon, \delta)$-differentially private if for all $S \subseteq 
	\mathcal{O}$ and every $x, y \in \mathcal{D}$ such that $x \sim y$:
	\[
	\Pr(\mathcal{A}(x) \in S) \leq e^{\varepsilon}\Pr(\mathcal{A}(y) \in 
	S)+\delta 
	\text{.}
	\]
\end{definition}

As Definition~\ref{def-diff-privacy} indicates, 
randomisation is an essential component of differential privacy. 
In fact, no deterministic algorithm satisfies this privacy notion. 

This paper addresses the problem of providing information about the 
importance of users in a social network in a differentially private manner. 
We consider that the ability 
of determining the status of users within their social network is useful, 
but we consider revealing the connectivity information 
between two users to be a privacy intrusion . For example, in the social network induced 
by the contact tracing data collected by governments during the COVID-19 
pandemic, determining high-centrality individuals helps health authorities 
to more efficiently allocate scarce resources such as testing capabilities. 
However, while conducting this study, individuals (especially those who test negative) 
must be protected from the risk of having their potentially sensitive 
contact information revealed. 

There exist various centrality measures to calculate the importance of a user 
in a social network. Yet, only degree centrality has been made
differentially private while preserving the node ordering induced 
by the noiseless measure~\cite{HLMJ09}, which is arguably the most relevant 
utility criterion for centrality measures. Considering that different centrality 
measures provide different insights on the role of nodes in a graph, 
it is of considerable interest to understand to what extend 
other centrality measures are amenable to differential privacy. 
\vspace{2mm}

\noindent \textbf{Contributions.} 
This article provides the first analysis of the \emph{eigenvector}, \emph{Laplacian} 
and \emph{closeness} graph centrality measures under the differential privacy model. 
An observation that has been made in previous work, 
albeit no proof has been provided, is that the use 
of the Laplace mechanism~\cite{DR2014} in most centrality 
measures leads to unbounded noise. Our analyses confirm such claim for the 
centrality measures under study, and we go 
further by characterising the noise incurred by the 
smooth sensitivity mechanism~\cite{NRS07}, which claims to 
reduce noise with respect to the Laplace mechanism. We provide lower and upper 
bounds on the noise needed to make each centrality measure differentially 
private based on the smooth sensitivity mechanism. We compare these bounds 
to the level of tolerable noise for each measure, defined as the amount of noise 
under which the node ordering induced by the centrality measure remains 
close to the original ordering with high probability. This comparison 
allows us to draw three 
different conclusions on the feasibility of using smooth sensitivity 
on a given centrality measure. If the tolerable noise is below the lower bound, 
then our results render accurate smooth sensitivity-based differential 
privacy infeasible. If the tolerable noise is above the upper bound, then the 
randomised algorithms described in this article can be used as effective and 
useful differentially private methods. Otherwise, if the tolerable noise is in 
between the lower and upper bound, we cannot draw any meaningful conclusion, 
except that further investigation is required. We empirically illustrate 
the trade-off between privacy and accuracy in real-life and synthetic 
social graphs for the three centrality measures. 
\vspace{2mm}

\noindent \textbf{Organisation.} The rest of this paper is organised as follows. 
We first review related work in Section~\ref{sec-related-work}. 
Then, we formally enunciate the scope of our study in Section~\ref{sec-formalization}. 
Theoretical results for eigenvector, Laplace and closeness centralities 
are given in Sections~\ref{sec-eigenvector}, \ref{sec-laplacian-centrality} 
and~\ref{sec-closeness}, respectively, whereas empirical results are presented 
in Section~\ref{sec-experiments}. Finally, we give our conclusions 
in Section~\ref{sec-conclusions}. 

\section{Related work}
\label{sec-related-work}

The \emph{Laplace mechanism}, which consists in adding noise drawn 
from a Laplace distribution with mean zero 
and variance $2 \left( \Delta f/\varepsilon \right)^2$, 
is the most common approach to satisfy 
differential privacy~\cite{DR2014}. In the previous formula, $f$ is the query 
function 
of interest, $\varepsilon$ the privacy parameter, and $\Delta f$ 
the \emph{global sensitivity} of $f$, 
defined as $\Delta f = \max_{x, y \in \mathcal{D}, x \sim y} \Vert f(x) - f(y) 
\Vert_1$. Because noise is proportional to global sensitivity, and many useful 
functions have large global sensitivity, Nissim et al. later introduced 
a sampling method based on \emph{smooth sensitivity}~\cite{NRS07}, 
which claims to reduce noise with respect to the Laplace mechanism. 

For graphs, the general notion of $(\varepsilon, \delta)$-differential privacy 
has been instantiated in several ways, each depending on the definition 
of the neighbouring dataset relation $\sim$. The most commonly used notion, 
\emph{$(\varepsilon, \delta)$-edge differential privacy}, states that two graph 
datasets are neighbouring if they differ in exactly one edge. 
In this case,  
differential privacy ensures that the output of the function does not leak 
information as to whether two users are connected. 
Alternatively, two graphs are considered as vertex-neighbouring if they differ 
by exactly one vertex, and all edges incident to that vertex. 
In this case, differential privacy ensures 
that the output of the algorithm does not leak information as to whether a user 
is in the graph. 
In this paper, we use a generalisation of $(\varepsilon, \delta)$-edge 
differential privacy for the case of edge-weighted graphs. 

Differentially private degree sequences~\cite{HLMJ09,KS12}, 
and the related notion of degree correlations~\cite{WW13}, 
were the earliest focus of research on the application 
of differential privacy to graph data. The degree sequence of a graph 
is a particularly good statistic in terms of amenability to differential privacy, 
especially under the notion of edge-differential privacy, 
as it requires to add very small amounts of noise to guarantee privacy. 
This has allowed studies to deepen on techniques for improving 
the final (post-processed) results. The general trend in publishing these statistics 
under DP consists in adding noise to the original sequences 
and then post-processing the perturbed sequences to enforce 
or restore certain properties, such as graphicality~\cite{KS12}, vertex order
in terms of degrees~\cite{HLMJ09}, etc. These studies are also particularly relevant 
from the perspective of this paper, as degree is the most straightforward 
vertex centrality measure. To the best of our knowledge, degree is in fact 
the only vertex centrality measure for which differentially private methods 
have been proposed, and the present paper is the first comprehensive analysis 
in this field. 

Computing degree sequences and degree correlations is often seen as an intermediate 
step in building graph generative models, from which synthetic graphs 
are later sampled 
and released to analysts~\cite{MW09,KS12,SZWZZ11,WW13,XCT14,JYC16,CMR2020}. 
Under this approach, differential privacy is applied in computing model parameters, 
and sampling is performed as post-processing, so the synthetic graphs 
preserve the same privacy guarantees as the models themselves. 
In addition to degree sequences and degree correlations, a differentially private 
version of the Kronecker graph model~\cite{LF07} is used in~\cite{MW09}, 
whereas the $dK$-graph model, which is based on differentially privately 
counting the occurrences of specific subgraphs with $K$ vertices 
(e.g. length-$K$ paths) is introduced in~\cite{SZWZZ11}. 
The \emph{hierarchical random graph} (HRG) 
model~\cite{CN08} was shown in~\cite{XCT14} to allow for further reductions 
of the amount of added noise. Furthermore, a differentially private version 
of the \emph{attributed graph model} (AGM)~\cite{PMFNG14} was introduced 
in~\cite{JYC16}, allowing to generate differentially private graphs 
featuring attributed nodes. This approach was extended in~\cite{CMR2020} 
to account for the community structure of the graph. 
The aforementioned generative model-based approaches have required 
to develop mechanisms for computing additional graph statistics under DP, 
e.g. community partitions~\cite{NIR16} and subgraph count queries 
such as the number of length-$K$ paths~\cite{SZWZZ11} 
and the number of triangles of either the entire graph~\cite{KRSY14,WWZX12,ZCPSX15} 
or that of a subgraph~\cite{CMR2020}. 

Recently, differentially private methods leveraging the randomized response strategy 
for publishing a graph's adjacency matrix were proposed in~\cite{ST2020}. 
Randomized response treats the adjacency matrix of the graph as a series 
of answers to the yes/no question ``\emph{are vertices $u$ and $v$ connected?}'', 
instead of numerical values. Thus, the randomisation is achieved 
by giving the true answer to this question with a given probability $p$, 
and a random answer with probability $1-p$. 

Differentially private computation methods for other graph problems, 
including vertex cover, set cover, min-cut and $k$-median, are described 
in~\cite{Gupta2010}. Despite the existence of the aforementioned results, 
it is important to highlight that the accurate, 
differentially private computation of very basic graph queries, 
e.g. graph diameter, has revealed to be infeasible or considerably challenging. 
This paper contributes several results of this type concerning centrality measures, 
as we show that there exist graph families for which a meaningful privacy protection 
in the computation of closeness, eigenvector and Laplacian centralities leads 
to arbitrarily inaccurate node rankings. 

\section{Privacy goal, notation and problem statement}
\label{sec-formalization}

This section introduces notation and definitions necessary to formalise the 
privacy problem we address. 
In particular, we define the type of data to be analysed, the information 
to be queried from data, the information we intend to protect 
and the privacy-preserving technique used to protect that information. 

\subsection{Domain of analysis: weighted social graphs}

The type of graphs we consider are weighted, connected and undirected. We use  
$G=(V,E, W)$ to denote one of such graphs, where $V$
is a set of vertices, $E$ a symmetric relation on the set of vertices 
representing edges, and 
$W: V \times V \rightarrow \mathbb{R}^{+}$ a total function mapping weights to 
edges. Because $G$ 
is undirected, we require $W$ to satisfy $W(u, v) = W(v, u)$ 
for every $(u, v) \in E$. We also require consistency between $E$ and $W$, 
in the sense that $W(u, v) > 0 \iff (u, v) \in E$.  
That is, edges in $E$ must feature non-zero weights, 
whereas the weight of a non-existing edge is considered to be zero by convention. 
We use $\graphuni$ to denote the universe of graphs of the type described above. 

\subsection{Information of interest: vertex centrality}

Vertex centrality measures score the structural importance of users 
within a social graph. In this paper, we treat a centrality measure as a function 
on (a~subset of) $\mathcal{G}$ yielding a vector 
$(y_1, \ldots, y_n) \in \mathbb{R}^n$, where $n$ is the order of the input graph 
(which corresponds to the number of users in the network). 
Formally, in order to account for graphs of different order, we define a centrality 
measure as a finite family of functions $\mathcal{C} = \{C_1, C_2, \ldots\}$ 
where, for every positive integer $i$, the domain of $C_i$ is 
$\graphuni_i = \{(V, E, W) \in \graphuni \mid |V| = i\}$ 
and its co-domain is $\mathbb{R}^i$. 
For the sake of simplicity, we assume that vertices in a graph satisfy an arbitrary, 
but fixed, total order. That is, given a graph $G = (V, E, W)$ 
it holds that $V$ is isomorphic to $\{1, \ldots, |V|\}$ under the total order $<$. 
Intuitively, given a graph $G$ of size $n$
and a centrality measure $C_n$, the vector $C_n(G) = (y_1, \ldots, y_n)$ assigns 
a score to each vertex in the network which quantifies its centrality. 
Considering $v_1 < v_2 < \ldots < v_n$ to be the totally ordered set of vertices 
of $G$, we say that $v_i$ is more important (or more central) than $v_j$ 
(for any pair $i,j \in \{1, \ldots, n\}$) if and only if $y_i > y_j$. 

\subsection{Privacy goal: $(\varepsilon, \delta)$-differential privacy} 

Although we allow the vector 
of centrality scores to be obtained from the social graph, we claim 
that information on the users relations, i.e. the weights of the connections 
among users in the graph, should remain private. This protects users from adversaries 
willing to learn how they interact with each other. For example, consider a 
social graph where the weight of each edge depends on the number of e-mails 
exchanged between the two connected users. Assuming the social graph belongs 
to a company $X$, it may not come as a surprise that top executives play 
central roles in that graph. Yet, revealing the number of e-mails exchanged 
by, for instance, the company manager with the rest of employees, 
may compromise the company's private operations 
or even leak the nature of the manager's personal relations with other employees. 

We consider in this article a generalisation of the popular 
edge-neighbouring 
relation~\cite{HLMJ09,KS12,WWZX12,KRSY14,ZCPSX15,NIR16,CMR2020}, which allows 
to reason about the protection of the weight of a connection rather than its 
mere existence. We call this neighbouring relation the \emph{$\theta$-edge 
neighbouring} and define it as follows. 

\begin{definition}[$\theta$-edge 
neighbouring]\label{def-theta-edge-neighbouring} 
	Given a positive real value $\theta$, the \emph{$\theta$-edge-neighbouring} 
	relation on $\graphuni$, denoted 
	$\thedgerel$, is the symmetric closure of the smallest relation satisfying, 
	for every $G = (V, E, W), G' = (V', E', W') \in \graphuni$
	\begin{align*}
	&G \thedgerel G' \iff V = V' \wedge \exists (u, v) \in E
	\colon 0 < \vert W(u, v) - W'(u, v)\vert \le \theta \wedge \\ 
	& \quad 
	\forall (x, y) 
	\in V \times V \colon W(x, y) = 
	W'(x, y) \iff \{x, y\} \neq \{u, v\}. 
	\end{align*}
\end{definition}

The neighbouring relation $\thedgerel$ can be defined algorithmically as follows: 
two graphs $G_1$ and $G_2$ are neighbouring if $G_1$ can be obtained from $G_2$ 
by increasing or decreasing the weight of one and only one edge in $G_2$ 
by up to a maximum value $\theta$. Therefore, a differentially private output 
with respect to $\thedgerel$ will guarantee that the adversary cannot differentiate 
the real graph from another one where the weight of a given edge differs 
by less than $\theta$. That is, the adversary can determine with sufficient 
certainty that the weight value lies in some interval, but cannot improve 
the granularity of this interval beyond a radius $\theta$ 
without sacrificing certainty. In particular, if $\theta$ equals the maximum weight 
of an edge, then differential privacy based on the notion 
of $\theta$-edge neighbouring datasets can be used to effectively prevent 
the adversary from learning the weight of any~edge. 
For ease of exposition and whenever it does not lead to confusion, we will use 
in the remainder of this article $\sim$ as 
a shorthand notation for $\thedgerel$. 

\subsection{Perturbation techniques: Laplace mechanism and smooth sensitivity}

In order to make a centrality measure $C_n$ 
differentially private, one needs to add noise to the outputs of $C_n$. 
The Laplace mechanism~\cite{DR2014} adds noise proportional to the difference 
between the outputs of $C_n$ on every pair of 
sufficiently close inputs. Such a difference is known as \emph{global 
sensitivity}. 
 
\begin{definition}[Global sensitivity] 
\label{def-sensitivity} 
Let $f:\graphuni \rightarrow\mathbb{R}^n$ be a deterministic function 
whose co-domain is the real coordinate space of $n$ dimensions. 
Global sensitivity with respect to $\sim$ and $f$ is defined as follows. 
$$\sensitivity{f}{\sim}=\max_{G,G'\in \graphuni, G\sim G'} \Vert f(G)-f(G') 
\Vert_1.$$ 
\end{definition} 

The second randomised mechanism that we employ in this article is based on 
a less stringent notion of sensitivity 
called \emph{smooth sensitivity}~\cite{NRS07,KRSY14}, 
which, rather than considering any pair 
of neighbouring graphs, depends on the raw graph 
and considers its neighbouring graphs.

\begin{definition}[$\beta$-smooth sensitivity]
\label{def-beta-smooth-sensitivity}
Let $f:\graphuni \rightarrow\mathbb{R}^n$ be a deterministic function and 
$\sim$ a neighbouring relation between 
graphs. Let $d\colon \graphuni \times \graphuni  \rightarrow \reals$ 
be a distance measure 
defined by $d(x,y) = k$ if $k$ is the smallest positive integer such that there 
exists $z_0, \ldots, z_k \in \graphuni$ satisfying that $z_0 = x$, $z_k = y$ 
and $z_0 \sim \cdots \sim z_k$. Given a real value $\beta$, the 
\emph{$\beta$-smooth sensitivity} of $f$ around $x \in \graphuni$ is 
	\[
	S^*_{f, \beta}(x) = \max_{y \in D^n} \left( \locsens{f}{\sim}{y} \cdot 
	e^{-\beta d(x, 
		y)} \right)
	\]
where $\locsens{f}{\sim}{y} = \max_{G'\in \graphuni, G \sim G'} \Vert 
C_n(G)-C_n(G') 
\Vert_1$ is known as the \emph{local sensitivity} of $y$ with respect to $f$ 
and $\sim$.
\end{definition}

Local and smooth sensitivity have been used as auxiliary tools 
in differentially private computations~\cite{KRSY14,ZCPSX15,CMR2020}. 
However, there does not exist yet a procedure to 
calculate or estimate these parameters with respect to standard 
centrality measures, such as eigenvector, Laplacian and closeness centralities. 
In what follows, we address that limitation by providing bounds on the global, 
local and smooth sensitivities of the three centrality measures under study. 

\section{Eigenvector centrality}
\label{sec-eigenvector}

Eigenvector centrality uses linear algebraic properties of the adjacency matrix 
of a graph to determine the influence of each node. The centrality score of a 
node $u$, denoted $e\mhyphen score_G(u)$, is calculated recursively as the 
weighted sum of the scores of its neighbours divided by a constant $\lambda$. 

\[
e \mhyphen score_{G}(u) = \frac{1}{\lambda}\sum_{(u, v) \in E} W(u, v) \times e \mhyphen score_{G}(v)
\]

Let $C^e_n$ denote the eigenvector centrality function with domain 
$\graphuni_n$ and co-domain $\mathbb{R}^n$ defined as 
$$C_e^n(G) = (e \mhyphen score_G(v_1), e \mhyphen score_G(v_2), \ldots, 
e \mhyphen score_G(v_n))$$ 
for every graph $G$ with totally ordered set of vertices $v_1 < \ldots < v_n$. 
Let $A$ be the weight (positive) matrix of $G$. Then we may write, 

\[
\lambda \times C^e_n(G) = A \times  C^e_n(G)
\]

This means that $C^e_n(G)$ is an eigenvector of $A$ with corresponding 
eigenvalue~$\lambda$. Note that for this centrality measure we are not 
interested in every eigenvector, but rather the one with the largest associated 
eigenvalue. To summarise, the eigenvector centrality function $C^e_n(G)$ of a 
graph $G$ is the eigenvector associated with the largest eigenvalue of the 
weight (positive) matrix of $G$. 

\subsection{Smooth sensitivity of $C^e_n(G)$}

We start by analysing the local sensitivity of $C^e_n(G)$. 

\begin{theorem}\label{theo-eigenvector-bound}
Let $G = (V, E, W)$ be a graph with $|V| = n$. Let $A$ be the weighted 
adjacency matrix of $G$ with eigenvalues $\lambda_1 \leq \ldots 
\leq \lambda_n$. 
The local sensitivity $\locsens{C^e_n}{\sim}{G}$ of $G$ with respect to 
the $\theta$-edge neighbouring relation 
$\sim$ and the eigenvector centrality measure $C^e_n$ is upper-bounded by

$$\locsens{C^e_n}{\sim}{G}
\leq \sqrt{n} \frac{2|\theta|}{ | \lambda_n - \lambda_{n-1}|}$$ 
\end{theorem}

\begin{proof}

Let $\hat{v_i}$ be 
the unitary vector of length $n$ with a $1$ in the $i$-th position and $v(A)$ the eigenvector of maximum eigenvalue in $A$. 
Consider a graph $G' = (V, E, W')$ satisfying that $G \sim G'$ and $G \neq G'$. Take two vertices in $G$, say $v_i$ and $v_j$, such that $W'(v_i, v_j) = W(v_i, v_j)+\varepsilon$ for some non-zero real value~$\varepsilon$. Note that such a pair of vertices ought to exist as $G \neq G'$. It follows that the adjacency matrix $A'$ of $G'$ is related to the adjacency matrix of $G$ by $A' = A + \varepsilon 
\hat{v_i}\cdot 
\hat{v_j}^{T}$. 
This means that, 

\[
\locsens{C^e_n}{\sim}{G} = \Vert C_n^e(G)-C_n^e(G') 
\Vert_1  = \Vert 
v(A) - v(A + \varepsilon 
\hat{v_i}\cdot 
\hat{v_j}^{T})
\Vert_1
\]

First, notice that the eigenvectors $v(A)$ and $v(A')$ satisfy  

\begin{equation}
\begin{aligned}
& \|v(A)-v(A')\|_2 \leq \sqrt{2} \sin \Theta(v(A),v(A'))
\end{aligned}
\end{equation}
where $\Theta(v(A),v(A'))$ denotes the angle between $v(A)$ and $v(A')$. 
Now, from Davis-Kahan~\cite{DK69} we have that 

\begin{equation}
\begin{aligned}
& \sin \Theta(v(A),v(A')) \leq \frac{2\| \varepsilon \hat{v_i} \hat{v_j}^T\|}{min_{n\neq j} | \lambda_{n} - \lambda_{j}|} \\
& \|v(A)-v(A')\|_2 \leq \frac{2\| \varepsilon \hat{v_i} \hat{v_j}^T\|}{min_{n\neq j} | \lambda_n - \lambda_j|}
\\
& \|v(A)-v(A')\|_2 \leq \frac{ 2|\varepsilon| }{min_{n\neq j} | \lambda_n - \lambda_j|}
\\
& \|v(A)-v(A')\|_2 \leq \frac{2 |\varepsilon| } {| \lambda_n - \lambda_{n-1}|}
\end{aligned}
\end{equation}

However, we are interested in $ \|v(A)-v(A')\|_1$, not  $\|v(A)-v(A')\|_2$. Therefore we have 

\begin{equation}
\begin{aligned}
& \|v(A)-v(A')\|_1 \leq  \sqrt{n} \|v(A)-v(A')\|_2 \leq \sqrt{n} \frac{2|\varepsilon|}{ | \lambda_n - \lambda_{n-1}|}
\end{aligned}
\end{equation}

Using the fact that $|\varepsilon| \leq |\theta|$ we conclude the proof.
\end{proof}

In the worst case, these bounds are tight. However, if one restricts the graph 
models under consideration, it is possible to obtain tighter 
bounds~\cite{EBW18}. Here we consider only the general case, and leave more 
restricted models for future work.  
Having bounded the local sensitivity, we would now like to use this to bound $\beta$-smooth sensitivity.
Unfortunately, for weighted graphs, $\beta$-smooth sensitivity can be arbitrarily large, 
as shown next.  

\begin{theorem}\label{theo-eigenvector-bound-smooth-weighted}
There exists a weighted connected graph $G$ such that its $\beta$-smooth sensitivity 
$S^*_{C^e_n, \beta}(G)$ is unbounded. 
\end{theorem} 

\begin{proof}

We proceed by construction of an example. We wish to bound
\begin{equation}S^*_{C^e_n, \beta}(G) = \max_{G' \in D^n} \left( 
\locsens{f}{\sim}{G'} \cdot 
	e^{-\beta d(G, 
		G')} \right)
\end{equation}

Therefore it suffices to show that there is some $G$, that has a nearby neighbour $G'$ with an arbitrarily bad spectral gap. 
An example of such graph $G$ is demonstrated in what follows. Consider a simple graph with two vertices $a,b$. Each vertex is linked to the other by an edge with weight 
$\theta+\epsilon$, where $\epsilon$ is some constant, which we set arbitrarily close to zero. In this case the eigenvalues of $G$'s adjacency matrix $A$ are $\lambda_1=\theta+\epsilon$, and $\lambda_2=-(\theta+\epsilon)$. After making a $\theta$ reduction to the edge weight we now have $\lambda_1=\epsilon$, and $\lambda_2=-(\epsilon)$, giving a spectral gap of $|2\epsilon|$. We thus have 

\begin{equation}S^*_{C^e_n, \beta}(G) \leq \max_{G' \in D^n} \left( 
 \frac{2\|\theta\|}{\epsilon} \cdot 
	e^{-\beta\theta} \right)
\end{equation}

Since $\epsilon$ can be made arbitrarily close to zero, the smooth 
sensitivity of $G$ becomes arbitrarily large. 
\end{proof}

Theorem~\ref{theo-eigenvector-bound-smooth-weighted} agrees with results previously determined by \cite{SegarraR16}, showing that if a graph has a bad spectral gap, then no useful centrality rankings can be computed, which agrees with our bound above. Another way of looking at this result is that as long as there is a graph in the neighbourhood of $G$ with a bad spectral gap we have to add enough noise to the graph to completely remove any utility from the centrality measure. 
This is analysed in detail in what follows.

\subsection{Impracticality of differential privacy for eigenvector centrality}
\label{sec-utility-loss-th}

We explain the impact on utility from adding noise proportional to the bound 
on local sensitivity given above. 
Since local sensitivity serves as a lower bound on smooth sensitivity, the fact 
that this amount of noise, which still guarantees no privacy, is already sufficient 
for destroying utility, means that $(\varepsilon,\delta)$-differentially private 
versions of this method are impractical. 

\begin{figure}[thb]
 \centering
\includegraphics[scale=.135]{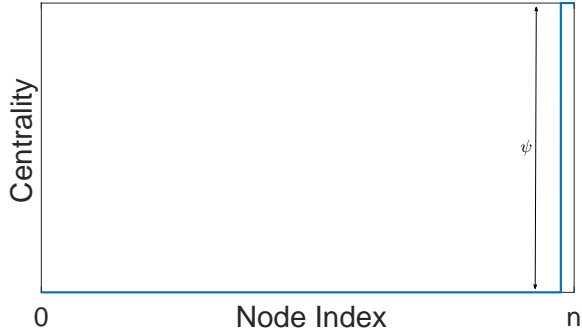}

\caption{Example of an ideal centrality ranking. The majority of the nodes 
have a low ranking, and a small $10\%$ minority have a high ranking. 
The gap separating the low ranked nodes and the high ranked nodes is $\psi$.}
\label{fig:IdealGraph}
\end{figure}

Consider the centrality ranking shown in Figure \ref{fig:IdealGraph}. Here we 
have a situation that is nearly optimal for centrality ranking algorithms. We 
have a small percentage of nodes that are very important, and a vast majority of 
nodes that are not important at all. The important nodes are separated from the 
unimportant nodes by a gap of $\psi$. Given the upper bounds 
for local 
sensitivity we have previously computed, we show that we will missclassify all 
the nodes with high probability, unless the gap $\psi$ is unrealistically 
large. Since in practice the distribution of node centrality will be much less 
optimally distributed, this means that after adding noise calibrated to 
smooth sensitivity we will, in general, preserve no~to little utility. 

In order to enforce privacy, we need to add a noise vector consisting of $n$ random 
variables from the Laplace distribution $L(0,b)$, where $b$ is the bound 
we previously computed for the local sensitivity. 
This means that our distribution has a variance of $2b^2$, and a mean $0$. 

\begin{theorem}\label{theo-chey}
In order for a diferentially private version of eigenvector centrality 
to correctly classify all but a constant number of nodes $c$, the gap $\psi$ 
between the important and unimportant nodes must be $O(\sqrt{n})$. 
\end{theorem}

\begin{proof}

In order to misclassify an unimportant node as an important node, we must add noise proportional to the gap $\psi$. We are computing the noise as a vector of $n$ i.i.d random variables drawn from a Laplace distribution $L(0,b)$, where $b=\frac{\sqrt{n}|\theta|}{\delta(A)}$, where $\delta(A)$ is the spectral gap of $A$. Since we are working with the Laplace distribution we can use the cumulative distribution function $1-F(x)=1-(1-\frac{e^{-x/b}}{2}), x \geq 0$ to compute the probability that an element of our noise vector is greater than $\psi$. Then, $n$ times this probability will be the number of misclassified vertices. We wish to compute the minimum size so that the number of misclassified vertices is less than some constant amount $c$. We~thus have 

\begin{equation}
\begin{aligned}
    (1 - P(\psi))n < c \\
    (1 - ( 1 - \frac{e^{-\frac{\psi \delta(A)}{\sqrt(n)|\theta|}}}{2}))n < c \\
    (e^{-\frac{\psi \delta(A)}{\sqrt(n)|\theta|}}))n < 2c \\
    (e^{-\frac{\psi \delta(A)}{\sqrt(n)|\theta|}}) < \frac{2c}{n} \\
    -\frac{\psi \delta(A)}{\sqrt(n)|\theta|} < \log(\frac{2c}{n}) \\
    - \psi < \frac{\log(\frac{2c}{n})\sqrt{n}|\theta|}{\delta(A)} \\
    \psi > \frac{-\log(\frac{2c}{n})\sqrt{n}|\theta|}{\delta(A)} = O(\sqrt{n})
\end{aligned}
\end{equation}

\end{proof}

A gap of $O(\sqrt(n))$ is grossly unrealistic even for graphs that have an extremely favourable centrality ranking. 
Since the bounds on local sensitivity for eigenvector centrality are tight for some graphs, in the worst case we cannot offer any computational utility after adding noise. We leave open the question of what the average case is for local sensitivity. However, since local sensitivity serves as a lower bound on smooth sensitivity, it seems likely that in practice the average case is not much better than the worst case. 
In Section~\ref{sec-experiments} we present empirical results 
which support this claim. 

\section{Laplacian centrality}
\label{sec-laplacian-centrality}

Let $G$ be an undirected and weighted graph with vertices $\{v_1, \ldots, 
v_n\}$ and weight matrix $W$. The \emph{Laplacian 
	matrix} of $G$ is defined by $L = 
X - W$, where $X$ is the diagonal matrix satisfying that $X_{ii} = \sum_{j = 
	1}^n 
W_{ij}$ for every $i \in \{1, \ldots, n\}$.

\newcommand{\laplacent}{C_{\mathcal{L}}^n}

The Laplace energy of $G$, denoted $\Lambda(G)$, 
is the sum of the squares of the eigenvalues of the \emph{Laplacian 
	matrix} of $G$, i.e. $\Lambda(G) = \lambda_1^2+  \cdots + \lambda_{n}^2$  
	where 
$\lambda_1, \cdots, \lambda_{n}$ are the eigenvalues of $G$. Considering 
$G_{v_i}$ to be the 
graph resulting from removing $v_i$ from $G$, 
the 
\emph{Laplacian centrality} vector of a graph $G$ is the vector 
$\laplacent(G) = 
(c_1, \ldots, c_n)$ where, 

\[
c_i = \frac{\Lambda(G)- \Lambda(G_i)}{\Lambda(G)} \ \forall i \in \{1, 
\ldots, 
n\} 
\text{.} 
\]

We wish to bound  $\locsens{\laplacent}{\sim}{G} = \Vert 
\laplacent(G)-\laplacent(G')\Vert$ , 
and then use this to bound $S^*_{\laplacent, \beta}$. We will make use of 
two well known theorems. The first is Weyl interlacing inequality theorem 
\cite{Weyl}. This theorem states that given two Hermitian matrices $A$, with 
eigenvalues $\lambda_i$, and a Hermitian perturbation matrix~$P$, one can 
define a new matrix $M=A+P$, and bound its eigenvalues. In particular, the 
eigenvalues $\mu_i$ of M are bounded by $|\mu_i-\lambda_i| \leq \|P\|$. 
Here the norm may be any consistent matrix norm, so we consider the $\|P\|_\infty$ 
norm. We note that if we make $k$ changes of size $\theta$ to $G$, then $\|P\|_\infty = k \theta$\textbf{}. 

We will also make use of a theorem from \cite{deletingedge}, which states that for an unweighted graph the eigenvalues of a graph created by deleting a vertex  $v_i$ are bounded as $\lambda_i-1 \leq \lambda_i^{v} \leq \lambda_{i+1}$, where $\lambda_i^{v}$ are the eigenvalues of the new graph Laplacian, and $\lambda$ are the eigenvalues of the original graph. We begin by extending this theorem to weighted graphs. In doing so we follow the same argument as the original theorem. 

\begin{theorem}\label{theo-horribletheorem}

Let $L$ be the Laplacian of a undirected, weighted graph, with maximum weighted edge $\delta$. Let $L^v$ be the graph created by deleting an edge. Let $P$ be the principal sub-matrix of $L$, created by deleting a row and column of $L$. Let the eigenvalues of $L$ be $\lambda_n ...\geq \lambda_i ...\geq \lambda_1$.  Let the eigenvalues of $L^v$ be $\lambda^{v}_n ...\geq \lambda^{v}_i ...\geq \lambda^{v}_i$. Finally let the eigenvalues of $P$ be $\rho_n ...\geq \rho_i ...\geq \rho_1$. Then we have $ \lambda_i-\delta \leq \lambda_i^{v}$.

\begin{proof}
Since our graph is undirected, we have that $L$ is a symmetric matrix, therefore by the Cauchy interlacing theorem, we have that $\lambda_1 \leq \rho_i ..\leq \rho_{n-1} \leq \lambda_n$. We now show that $\rho_i \leq \lambda^v_{i} + \delta$. Define $L_v=P-L^v$. $I_v$ is a diagonal matrix wiht all zero entries except for the $j$-th diagonal entry, which is at most $\delta$, iff $j$ is conected to $v$ in $G$, with weight $\delta$. We iterate over $\forall i \in {1,...,n-1}$. By the Courant-Fischer Theorem \cite{deletingedge} we have

\begin{equation}
\begin{aligned}
    \rho_{n-i+1}=\max_{U} \min_{x\in U} { \frac{x^tPx}{x^tx} : U \subseteq \reals^n, \dim(U)=i, x \in U = span(U)	} \\
     \rho_{n-i+1}=\max_{U} \min_{x\in U} { \frac{x^t(L^v+I_v)x}{x^tx} : U \subseteq \reals^n, \dim(U)=i, x \in U = span(U)	} \\
      \rho_{n-i+1} \leq \max_{U} \min_{x\in U} { \frac{x^t(L^v)x}{x^tx} : U \subseteq \reals^n, \dim(U)=i, x \in U = span(U)	} + \\
      \max_{U} \min_{x\in U} { \frac{x^t(I_v)x}{x^tx} : U \subseteq \reals^n, \dim(U)=i, x \in U = span(U)	} \\
      \leq \lambda^v_{n-i+1}+\delta
\end{aligned}
\end{equation}

We already have that $\lambda_1 \leq \rho_i ..\leq \rho_{n-1} \leq \lambda_n$. Substitute $\rho_{n-i+1} \leq \lambda^v_{n-i+1}+\delta$ and we are left with $\lambda_i \leq \lambda^v_{i}+\delta$, implying $\lambda_i -\delta\leq \lambda^v_{i}$.

\end{proof} 
\end{theorem}

\begin{theorem}\label{theo-giganticbounds}

Let $G$ be a weighted graph with Laplacian $L$,  eigenvalues $\lambda_n ...\geq 
\lambda_i ...\geq \lambda_1$, and max weight $\delta$.  Let $L^v$ be the 
Laplacian of the graph created by deleting some vertex $v$ from $G$, with 
eigenvalues $\lambda^v_n ...\geq \lambda^v_i ...\geq \lambda^v_1$. We have that 
$\locsens{\laplacent}{\sim}{G} = \Vert \laplacent(G)-\laplacent(G')\Vert \leq 
n\sum^n_{i=0} 
2(\lambda_i\delta)-\delta^2$. 

\begin{proof}
We have $c_i = \sum^{n}_{i=0} \lambda_i^2 - \sum^{n}_{i=0} \lambda^{v2}_i$. By Theorem \ref{theo-horribletheorem} we have that $\lambda^{v}_i \leq \lambda_{i-1}-\delta$, so $c_i \leq \sum^{n}_{i=0} \lambda_i^2 - \sum^{n}_{i=0} (\lambda_i -\delta)^2 =  \sum^n_{i=0}( 2(\lambda_i\delta)-\delta^2)$. Therefore $ \Vert C_n^e(G)-C_n^e(G')\Vert \leq n\sum^n_{i=0} (2(\lambda_i\delta)-\delta^2)$.
\end{proof}

\end{theorem}

We now need to bound $ S^*_{\laplacent, \beta} $. This reduces to an 
optimization problem.

\begin{theorem}\label{theo-giganticbounds2}
Let $G$ be a weighted graph with Laplacian $L$,  eigenvalues $\lambda_n ...\geq \lambda_i ...\geq \lambda_1$, and max weight $\delta$.  Let $L^v$ be the Laplacian of the graph created by deleting some vertex $v$ from $G$, with eigenvalues $\lambda^v_n ...\geq \lambda^v_i ...\geq \lambda^v_1$. We have that 
$ S^*_{\laplacent, \beta}(G) = \max_{G' \in D^n} \left( 
\locsens{\laplacent}{\sim}{G'} \cdot 
	e^{-\beta d(G, 
		G')} \right) = \Vert \laplacent(G)-\laplacent(G')\Vert e^{-\beta 
		\theta} 
		\leq  n\sum^n_{i=0} (2((\lambda_i+\theta)\delta)-\delta^2) e^{-\beta 
		\theta}$.
\end{theorem}

\begin{proof}
We know by Wely's theorem~\cite{Weyl} that any perturbation $P$ of the original graph $G$, will perturb the eigenvalues of the Laplacian $L$ of $G$ at most $\|P\|$. If~we take the worst case where the perturbation is tight then we have the bound above. If the perturbation is not tight, than the maximum will be reached at a higher $\theta$, and will thus be smaller. 
\end{proof}

For a given value of $\beta$ we can obtain the worst case $\theta$ by differentiating $n\sum^n_{i=0} (2((\lambda_i+\theta)\delta)-\delta^2) e^{-\beta \theta}=0$, with respect to $\theta$, then solving for $\theta$.  
Unfortunately, it can be seen that these bounds are likely to be very loose in 
practice, given their dependence on $n$, and the sum of the entire spectrum. 
Experimentally, we find that using these bounds to generate noise destroys all 
the information about the node ordering that we were seeking to preserve 
(see Section~\ref{sec-experiments}).

\section{Closeness centrality}
\label{sec-closeness}

Closeness centrality measures how well-connected a user is in a social network. Intuitively, the smaller the sum of the weights along a path between vertices $u$ and $v$, the better connected they are. It is worth noting that in this interpretation of connectivity the weights represent the cost of traversing an edge. We use $paths_G(u, v)$ to denote the set of all paths between $u$ and $v$ in a graph $G$. The connectivity score between $u$ and $v$ in $G$, denoted $S_G(u, v)$, is thus defined by 

\[
S_G(u, v) =  
\begin{cases}
\infty & \text{if $paths_G(u, v) = \emptyset$} \\
\min_{(x_1, \ldots, x_n) \in paths(u,v)} \sum_{i = 1}^{n-1} W(x_i, x_{i+1}) & 
\text{otherwise}
\end{cases}
\]

Note that a connectivity score is strictly larger than zero, while it could be 
$\infty$ if the graph is disconnected. To express the intuition that a low  
connectivity score makes a vertex more accessible, the closeness centrality 
score of a vertex $u$ in a graph $G$, denoted $c \mhyphen score_G(u)$, is 
defined by 

\[
c \mhyphen score_G(u) = \frac{1}{\sum_{v \in V} S_G(u, v)}
\]

\noindent
where, by convention, we take $1/\infty=0$. 

We use $C^c_n$ to denote the closeness centrality function with domain 
$\graphuni_n$ and co-domain $\mathbb{R}^n$ defined, for every graph $G$ with 
totally ordered set of vertices $v_1 < \cdots < v_n$, by

$$C^c_n(G) = (c \mhyphen score_G(v_1), c \mhyphen score_G(v_2), \ldots, c \mhyphen score_G(v_n)).$$

\begin{theorem}\label{theo-closeness-bound}
Let $G = (V, E, W)$ be a graph with $|V| = n$. The local 
sensitivity of $G$ with respect to the $\theta$-edge neighbouring relation 
$\sim$ and the closeness centrality 
measure $C^c_n$ is upper-bounded by

$$\frac{n\times (n-1)\times \theta}{\sum_{e \in E} W(e) \times (\sum_{e \in E} 
W(e) - \theta)}.$$ 
\end{theorem}

\begin{proof}
Let $G'$ be a graph such that $G \sim G'$, with $G = (V, E, W)$ 
and $G' = (V, E, W')$. Consider two vertices $u,v\in V$, 
and let the vertex sequence $x_1, x_2, \ldots,$ $x_p$ with $u = x_1$ and $v = 
x_p$ be a shortest $uv$-path 
in $G$. Similarly, let $x_1', x_2', \ldots, x_q'$ with $u = x_1'$ and $v = 
x_p'$  be a shortest $uv$-path in 
$G'$. 
It follows that 

\[
S_{G'}(u, v) = \sum_{i = 1}^{q-1} W'(x_i', x_{i+1}') \leq \sum_{i = 1}^{p-1} W'(x_i, x_{i+1}).
\]

This means that 

\[
S_{G'}(u, v) - S_G(u, v) \leq \sum_{i = 1}^{p-1} W'(x_i, x_{i+1}) - \sum_{i = 1}^{p-1} W(x_i, x_{i+1}). 
\]

Now, because $G \sim G'$, we obtain that $$\left|\sum_{i = 1}^{p-1} W'(x_i, 
x_{i+1})-\sum_{i = 1}^{p-1} W(x_i, x_{i+1})\right| \leq \theta,$$ as the weight of 
at most one edge is different, up to a maximum difference of~$\theta$, 
and by definition a shortest path does not go through the same edge twice. 
Hence $|S_G'(u, v) - S_{G}(u, v)| \leq \theta$ 
This gives the following bound: 

\[
\sum_{v \in V} |S_G'(u, v) - S_{G}(u, v)| \leq (n-1)\times \theta.
\] 

By the triangle inequality, we also obtain 

\[
\left | \sum_{v \in V} S_{G'}(u) - \sum_{v \in V} S_{G}(u) \right | \leq 
\sum_{v \in V} |S_{G'}(u, v) - S_{G}(u, v)| \leq (n-1)\times \theta.
\] 

The following sequence of algebraic development is useful to bound $\Vert C_n^C(G)-C_n^C(G') \Vert_1$: 

\begin{align*}
& \Vert C_n^C(G)-C_n^C(G') \Vert_1 = \sum_{u \in V} \left |c \mhyphen score_{G}(u) - c \mhyphen score_{G'}(u) \right | = \\
& \sum_{u \in V} \left | \frac{1}{S_G(u)} - \frac{1}{S_{G'}(u)} \right | = \sum_{u \in V} \frac{\left | S_{G'}(u) - S_G(u) \right |}{S_{G'}(u) \times S_G(u)} \leq \sum_{u \in V} \frac{(n-1)\times \theta}{S_{G'}(u) \times S_G(u)}.
\end{align*}

Now, we use the fact that $S_G(u) \geq \sum_{e \in E} W(e)$ and $S_{G'}(u) \geq \sum_{e \in E} W'(e)$ to obtain 

\[
\Vert C_n^c(G)-C_n^c(G') \Vert_1 \leq  \frac{n\times (n-1)\times 
\theta}{\sum_{e \in E} W'(e) \times \sum_{e \in E} W(e)}.
\]

Finally, the result follows from the fact 
that $$\sum_{e \in E} W'(e) \geq \sum_{e \in E} W(e) -\theta.$$ 
\end{proof}

It is worth remarking that the bound above is tight, as the equality holds 
for any complete graph of order two such that the weight of its sole edge 
is greater than $\theta$. 
We now turn to determining the bounds on smooth sensitivity.

\begin{theorem}\label{theo-smooth-bound}
	
	Let $G$ be a graph in $\graphuni_n$, $w$ the sum of its weights, and 
	$\thedgerel$ the $\theta$-edge neighbouring relation with $\theta > 0$. For 
	every positive real $\beta$, it 
	holds that 
	\[
	S^*_{C^c_n, \beta}(G) \leq \max_{k \in \mathbb{N^+}} \left( \frac{n\times 
		(n-1)\times \theta}{\left( 
		w - k\theta \right) \times (
		w - k\theta - \theta)} \cdot 
	e^{-\beta k} \right)
	\]
\end{theorem}
\begin{proof}
	Take a graph $G'$ whose distance to $G$ with respect to $\thedgerel$ is 
	$k$. It follows that, if $G = (V, E, W)$, then $G' = (V, E, W')$ for some 
	weight function $W' \neq W$. 
	From Theorem~\ref{theo-closeness-bound} we obtain that 
	
	\begin{equation}\label{eq-upper-bound-local}
\locsens{C_n^c}{\thedgerel}{G'} \leq \frac{n\times (n-1)\times 
	\theta}{\sum_{e \in 
		E} 
	W'(e) \times (\sum_{e \in E} 
W'(e) - \theta)}.
	\end{equation}

Now, recall from Definition~\ref{def-beta-smooth-sensitivity} 
that $d(G,G') = k$ means that $k$ is the smallest positive 
integer such that there exists $G_0, \ldots, G_k \in \graphuni$ 
such that $G_0 = G$, $G_k = G'$ and $G_0 \thedgerel \cdots 
\thedgerel G_k$. Therefore, it follows that $\sum_{e \in E} W(e)  - k 
\theta 
\leq 
\sum_{e \in E} W'(e)$. By substituting in Equation~\ref{eq-upper-bound-local} 
we obtain that, if $\sum_{e \in E} 
W(e) - k\theta - \theta > 0$, then, 

$$\locsens{C_n^c}{\thedgerel}{G'} \leq \frac{n\times (n-1)\times \theta}{\left( 
\sum_{e \in E} 
	W(e) - k\theta \right) \times (\sum_{e \in E} 
	W(e) - k\theta - \theta)}.$$ 

The proof is completed by using the inequality above and the definition 
of smooth sensitivity.
\end{proof}

Theorem~\ref{theo-smooth-bound} serves as the basis of an algorithmic approach 
to calculate a bound for smooth sensitivity, 
which is depicted in Algorithm~\ref{alg-smooth-close}.

\begin{algorithm}
	\caption{Given a graph $G$ of order $n$, and real values $\theta$, $\varepsilon$ 
	and $\beta$, outputs a randomised algorithm that satisfies $\left( \varepsilon, 
	\frac{1}{e^{\frac{\varepsilon}{\sqrt{n}\beta}}} \right)$-differential privacy, 
	if it exists.} 
	\label{alg-smooth-close}
	\label{alg-closeness}
	\begin{algorithmic}[1]
		\State Let $w$ be the sum of the edge weights in $G$ with $w > \theta$
		\State Let $f(x) = \frac{n\times 
	(n-1)\times \theta}{\left( 
	w - x\theta \right) \times (w - x\theta - \theta)} \cdot 
e^{-\beta x}$
		\State Let $k = 0$
		\While{$f(k) < f(k+1)$}
			\If{$w - 
			k\theta  
				- \theta \leq 0$}
			\State \Return Smooth sensitivity is undefined within $\mathbb{N}$.
			\EndIf
			\State $k = k+1$
		\EndWhile  
		\State Let 
		$\alpha = \frac{\varepsilon}{2}$ 
		\State Let $Z$ be random variable sampled 
		according to $h(z) = \frac{1}{2}\cdot e^{-|z|_1}$.
		\State \Return Algorithm 
		$\mathcal{A}(x) = C^c_n(x) + \frac{f(k)}{\alpha} \cdot Z$ 
		is $\left(\varepsilon, 
		\frac{1}{e^{\frac{\varepsilon}{\sqrt{n}\beta}}}\right)$-differentially private. 
	\end{algorithmic}
\end{algorithm}

Correctness of the algorithm above follows from Lemma~2.5, Example~3 in~\cite{NRS07} 
and the claim that the function $f(k) = \frac{n\times 
	(n-1)\times \theta}{\left( 
	w - k\theta \right) \times (
	w - k\theta - \theta)} \cdot 
e^{-\beta k}$ has a single maximum within the natural numbers domain. 

\section{Empirical analysis}
\label{sec-experiments}

Here we compare the effect on utility of adding noise calibrated 
to the exact value of local sensitivity, the bounds on local 
sensitivity enunciated above, and the smooth sensitivity (where possible) 
for a collection of real-life and synthetic social graphs. The results confirm 
our theoretical results that the 
noise needed to provide $(\varepsilon,\delta)$-differential 
privacy has the potential to severely affect utility. 

The graphs we have selected are SNAP/p2p-Gnutella05,  Wiki-Vote, a synthetic 
preferential attachment graph, and a synthetic small world graph, created using 
the Matlab CONTEST toolbox, using $\mathcal{O}(10^4)$ nodes and the default attachment 
settings. We generate these graphs one time, and then reuse them throughout our experiments, while 
the real-life social network graphs can be obtained from the 
SuiteSparse Matrix Collection~\cite{Davis2011}. 
For each graph, we computed the real rankings for the most important $5$, $10$ and $15$ 
percent of the nodes, and compare this to the differentially private rankings 
in~order to compute the precision of our method. 
For our experiments, we used $\varepsilon = 2$ and a value of $\theta$ 
equal to the maximum edge weight of the 
graph under analysis. Certainly, those values are the edge of the maximum 
privacy that can be offered via differential-privacy, yet they are plausible 
values for a practical setting. We leave for 
future work the full analysis of the privacy-utility trade-off 
over the entire domains of $\varepsilon$ and $\theta$. 

\begin{table}[!ht]
\begin{center}
 \begin{tabular}{c|ccc|ccc|ccc} 
 \multirow{2}{*}{\backslashbox{Graph}{Precision}} 
 & \multicolumn{3}{c|}{Local Sensitivity} & 
 \multicolumn{3}{c|}{B. L. Sensitivity} & \multicolumn{3}{c}{B. S. Sensitivity}\\ 
 \cline{2-10} 
 &5\%&10\%&15\%&5\%&10\%&15\%&5\%&10\%&15\%\\
 \hline
 p2p-Gnutella05 & $9$ & $13$ & $18$ & $5$ & $9$ & $15$ & \multicolumn{3}{c}{N/A}\\ 
 Wiki-Vote & $72$ & $67$ & $65$ & $6$ & $11$ & $15$ & \multicolumn{3}{c}{N/A}\\
 Synt. Pref. Attach. & $5$ & $9$ & $14$ & $6$ & $10$ & $15$ & 
 \multicolumn{3}{c}{N/A}\\
 Synt. Small World & $4$ & $10$ & $13$ & $5$ & $10$ & $15$ & \multicolumn{3}{c}{N/A}\\
 \hline
\end{tabular}
\end{center}
\caption{Precision values, in percentages, for eigenvector centrality 
at the top $5\%$, $10\%$ 
and $15\%$ of the rankings. Since we have no mechanism to calculate or 
effectively bound smooth sensitivity, the corresponding columns of the table 
are empty.}
\label{table1}
\end{table}

We begin our analysis by computing the 
percentage of correctly identified nodes in the 
top $5/10/15$ percent of the nodes in the perturbed graph using eigenvector 
centrality. The results are displayed in Table~\ref{table1}. They show 
that very few of the important nodes are preserved 
after perturbation. Although the noise added to the Wiki-Vote graph 
based on the local sensitivity parameter does not completely destroy the rankings,  
the precision 
quickly degrades when using the bound of the local sensitivity. The precision 
on the other graphs under consideration does not significantly differ from 
random noise. 

In Table~\ref{table2} we display the experimental results for closeness 
centrality. In~this case, even for local sensitivity, the outputs are dominated 
by random noise. 

\begin{table}[!ht]
\begin{center}
 \begin{tabular}{c|ccc|ccc|ccc} 
 \multirow{2}{*}{\backslashbox{Graph}{Precision}} 
 & \multicolumn{3}{c|}{Local Sensitivity} & 
 \multicolumn{3}{c|}{B. L. Sensitivity} & \multicolumn{3}{c}{B. S. Sensitivity}\\ 
 \cline{2-10} 
 &5\%&10\%&15\%&5\%&10\%&15\%&5\%&10\%&15\%\\
 \hline
 p2p-Gnutella05 & $5$ & $8$ & $15$ & $4$ & $9$ & $14$ & $4$ & $9$ & $4$\\ 
 Wiki-Vote & $5$ & $8$ & $15$ & $3$ & $9$ & $14$ & $3$ & $9$ & $14$\\
 Synt. Pref. Attach. & $4.5$ & $11$ & $15$ & $3.5$ & $10$ & $16$ & $3.5$ & $10$ & $16$\\
 Synt. Small World & $5$ & $10$ & $13$ & $5$ & $8$ & $15$ & $5$ & $8$ & $15$\\
 \hline
\end{tabular}
\end{center}
\caption{Precision values, in percentages, for closeness centrality 
at the top $5\%$, $10\%$ and $15\%$ of the rankings.}
\label{table2}
\end{table}

Finally, we examine the results for Laplacian centrality in Table~\ref{table3}. 
Unlike eigenvector and closeness centralities, Laplacian centrality does 
preserve a statistically significant number of the original nodes in the 
original ranking when applying noise proportional to local sensitivity. 
Unfortunately, because the bounds on local and smooth sensitivity for this measure 
are very loose, we are not able to achieve an accurate 
differentially private computation in practice. 

\begin{table}[!ht]
\begin{center}
 \begin{tabular}{c|ccc|ccc|ccc} 
 \multirow{2}{*}{\backslashbox{Graph}{Precision}} 
 & \multicolumn{3}{c|}{Local Sensitivity} & 
 \multicolumn{3}{c|}{B. L. Sensitivity} & \multicolumn{3}{c}{B. S. Sensitivity}\\ 
 \cline{2-10} 
 &5\%&10\%&15\%&5\%&10\%&15\%&5\%&10\%&15\%\\
 \hline
 p2p-Gnutella05 & $34$ & $36$ & $38$ & $6$ & $10$ & $15$ & $6$ & $10$ & $15$\\ 
 Wiki-Vote & $91$ & $81$ & $81$ & $6$ & $9$ & $16$ &  $6$ & $9$ & $16$\\
 Synt. Pref. Attach. & $89$ & $94$ & $90$ & $5$ & $9$ & $15$ & $5$ & $9$ & $15$\\
 Synt. Small World & $8$ & $13$ & $21$ & $5$ & $10$ & $15$ & $5$ & $10$ & $15$\\
 \hline
\end{tabular}
\end{center}
\caption{Precision values, in percentages, for Laplacian centrality 
at the top $5\%$, $10\%$ and $15\%$ of the rankings.}
\label{table3}
\end{table}

\section{Conclusions}
\label{sec-conclusions} 

We have presented the first analysis of the differentially private computation 
of non trivial centrality measures on weighted graphs. Our study covered 
eigenvector, Laplacian and closeness centralities. We presented lower 
and upper bounds on the amount of noise that randomised algorithms 
based on the smooth sensitivity approach need to add in order to make 
the output of each measure differentially private. Our results entail 
that differentially private computation of the three centrality measures 
via the smooth sensitivity approach is either infeasible, in the sense that 
there are large families of graphs for which smooth sensitivity is unbounded; 
or impractical, in the sense that even for the cases where smooth sensitivity 
is bounded, the required amounts of noise result in unacceptably large 
utility losses in terms of the quality of centrality-based node rankings.  
\vspace{2mm}

\noindent
\textbf{Acknowledgements.} The work of Yunior Ram\'irez-Cruz 
was funded by Luxembourg's Fonds National de la Recherche (FNR),
grant C17/IS/11685812 (\mbox{PrivDA}). Part of this article was written 
while Yunior Ram\'irez-Cruz was visiting the School of Information Technology 
at Deakin University.

\end{document}